\documentclass[aip,jcp,twocolumn,showpacs,superscriptaddress]{revtex4}

\usepackage{epsfig}
\usepackage{graphics}
\usepackage{bm}

\begin{document}


\title{Anomalous melting behavior of solid hydrogen at high pressures}

\author{Hanyu Liu}
\affiliation{State Key Laboratory of Superhard Materials, Jilin University, 130012 Changchun, China}
\author{E.~R.~Hern\'{a}ndez}
\affiliation{Instituto de Ciencia de Materiales de Madrid (ICMM-CSIC),
Campus de Cantoblanco, 28049, Madrid, Spain}
\author{Jun Yan}
\affiliation{LCP Institute of Applied Physics and Computational Mathematics -- Beijing 100088 Beijing, China}
\author{Yanming Ma}
\email{mym@jlu.edu.cn}
\affiliation{State Key Laboratory of Superhard Materials, Jilin University, 130012 Changchun, China}
\date{\today}

\begin{abstract}
Hydrogen is the most abundant element in the universe, and its properties under conditions of high temperature
and pressure are crucial to understand the interior of large gaseous planets and other astrophysical bodies. At
ultra-high pressures solid hydrogen has been predicted to transform into a quantum fluid, because of its high
zero-point motion. Here we report first-principles two-phase coexistence and Z-method determinations
of the melting line of solid hydrogen in a pressure range spanning from 30 to 600~GPa. Our results suggest that
the melting line of solid hydrogen, as derived from classical molecular dynamics simulations,
 reaches a minimum of 367~K at $\sim$430~GPa; at higher pressures
the melting line of the atomic Cs-IV phase regains a positive slope.
In view of the possible importance of quantum effects in hydrogen at such low temperatures, we also
determined the melting temperature of the atomic Cs-IV phase at pressures of 400, 500 and 600~GPa, employing
Feynman path integral simulations. These result in a downward shift of the classical melting line by $\sim$100~K,
and hint at a possible secondary maximum in the melting line in the region between 500 and 600~GPa, testifying
to the importance of quantum effects in this system.
Combined, our results  imply that the stability field of
the zero-temperature quantum liquid phase, if it exists at all, would only occur at higher pressures than previously thought.
\end{abstract}


\maketitle

The characterization of the physico-chemical properties of hydrogen at high pressures and temperatures is of
fundamental interest in physics, astrophysics, chemistry and planetary science. Previous theoretical
studies~\cite{Brovman:1972,Ashcroft:2000p300} have predicted that at sufficiently
high pressures solid hydrogen would transform into a quantum liquid phase that would constitute
a new state of matter, due to the fact that  hydrogen has a very
large zero-point energy at ultra-high pressures.
Results from first-principles molecular dynamics (MD) simulations have indicated that liquid hydrogen undergoes a
molecular to atomic transformation~\cite{Scandolo:2003p299}, which is predicted to occur at a pressure of
125$\pm$10~GPa along the 1500~K isotherm. Another study, employing similar techniques, has reported the
finding of reentrant behavior~\cite{Bonev:2004p296} in the melting of hydrogen, i.e. the existence of a maximum
along the melting line, predicted to occur at $\sim$80~GPa and $\sim$900~K, followed by a decrease of
the melting temperature at higher pressures. This prediction has been subsequently confirmed by
experiments~\cite{Deemyad:2008p505,Eremets:2009p293,Subramanian:2011}, which place the maximum at
$\sim$106~GPa and 1050$\pm$60~K. The occurrence
of reentrant melting behavior is understood to be a necessary, though by itself not sufficient, condition for the
existence of the high pressure quantum liquid phase predicted by Ashcroft~\cite{Ashcroft:2000p300}, and
thus it was taken as a strong indication of its existence by Bonev and coworkers~\cite{Bonev:2004p296}.
With the assumption that the negative slope of the melting line persists to high enough pressures,
it was estimated~\cite{Bonev:2004p296} that the quantum fluid state would occur at pressures close to 400~GPa.
Recent first-principles MD and quantum Monte Carlo~\cite{Morales:2009p323} simulations
have confirmed once more the reentrant behavior of the melting line, and determined a coexistence point
at which both the molecular and atomic fluids coexist with the solid (phase I); this coexistence point could either
be a triple point or, if the insulator-metal transition occurring in the liquid actually extends into the solid,
a quadruple point.

Here we report first principles calculations of the melting temperatures of five different crystalline phases
of solid hydrogen (four molecular ones and one atomic), which are the energetically most competitive phases in the
pressure range 30-600~GPa. These phases are the molecular {\em d-hcp} (phase I), partially ordered {\em hcp}
({\em po-hcp} or phase IV), {\em Cmca-4, Cmca-12\/} and
quasi-molecular {\em mC24\/} phases, plus the atomic {\em Cs\/}-IV phase. We have used
two reliable methods (two-phase coexistence~\cite{Morris:1994}
and the Z-method~\cite{Belonoshko:2006p287}) to predict the melting curve of hydrogen in this range of pressures. Our results
indicate that the high-pressure Cs-IV phase of hydrogen has a melting line with a positive slope. When considered
together with the reentrant behavior of the melting line at lower pressures, this results in the melting line
having a minimum value of 367~K at a pressure of $\sim$432~GPa. While the occurrence of this minimum along
the melting line does not preclude the existence of the predicted quantum liquid phase~\cite{Ashcroft:2000p300},
it would nevertheless push its stability field to higher pressures than previously thought.
However, these results are
based on classical simulations, and due to the lightness of hydrogen and the relatively low melting temperatures that
result in this range of pressures, it is necessary to consider the possible influence of quantum effects. This is a point
to which we return below.

\begin{figure}[!t]
\begin{center}
\epsfxsize=8cm
\epsffile{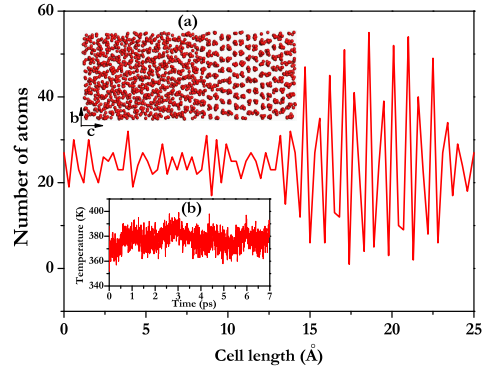}
\end{center}
\caption{(Color online) The main panel shows the particle density of the {\em Cs\/}--IV phase in coexistence
with liquid hydrogen at 500~GPa. Regularly spaced density peaks reveal the presence of crystal planes on the
right, whereas the small density oscillations around a constant value seen on the left are characteristic of the liquid phase.
The upper inset, (a), shows an instantaneous configuration resulting from the same coexistence simulation. Crystal
planes are clearly visible on the right side of the image, and the interface and liquid are discernible on the
left. The lower inset, (b), shows the time dependence of the instantaneous temperature of the simulation shown in
the main panel. The temperature has no tendency to drift, oscillating around an average value of 378~K with
a standard deviation smaller than 20~K, as obtained from the last 7~ps of simulation.}
\label{fig:coexistence}
\end{figure}

The two-phase coexistence approach consists of placing within the same simulation box the solid and liquid
phases directly in contact through an interface. The melting temperature is determined as that at which both
phases are seen to remain stably in coexistence.
At the first-principles level, such simulations are computationally
demanding, as they require large cells and long simulation times, but nevertheless they are accessible with
modern supercomputer facilities as recently demonstrated for the cases of Li~\cite{Hernandez:2010p562}
and Fe~\cite{Alfe:2009}.
In this work we have used first principles molecular dynamics simulations based on
density functional theory~(DFT). For this task we employed the Vienna {\em ab initio\/} Simulation
Package (VASP)~\cite{vasp}. Simulations were conducted in the NPH~ensemble (constant number of particles, N,
constant pressure, P, and constant enthalpy, H) following the algorithm of Souza and Martins~\cite{souza97}
as recently implemented in VASP~\cite{hernandez01,Hernandez:2010p562}. At a given
external pressure, the temperature adjusts spontaneously and tends towards the coexistence temperature, as the
unstable phase in the simulation is gradually consumed in favor of the stable one. By suitably repeating such
simulations at various temperatures it is possible to obtain the coexistence temperature as the temperature at which both phases stably
coexist and no drift in the instantaneous temperature is observed. The presence of both solid and liquid phases within
the simulation box can be easily corroborated either by direct inspection or by plotting the particle density
calculated at  a series of planes parallel to the interface [see Fig~(\ref{fig:coexistence})]. Crystal planes appear
as spikes of high density, whereas no such spikes are present in the liquid region. The simulation was constrained
to retain a tetragonal shape, with the two short sides (parallel to the plane of the interface) begin of equal
length. System sizes included 980, 960 and 960 hydrogen molecules and 2048 hydrogen atoms in the
($7\times 7 \times 10, 2\times2\times 5, 8 \times 8\times 8, 6 \times 4 \times 10$) simulation cells of the
{\em d-hcp\/}, {\em po-hcp\/}, {\em Cmca-4\/} and Cs-IV phases, respectively. Initial configurations
containing approximately equal amounts of the solid and liquid phases, plus the interface, were then generated.
When both phases were found to coexist stably for a minimum of 5~ps they were assumed to be in
equilibrium at the temperature and pressure conditions of the simulation. In these simulations the
Brillouin zone was sampled with  a grid of $2\times 2\times 1$~{\em k\/}-points; the self-consistency criterion
required on the total energy was a variation smaller than $2\times 10^{-5}$~eV between two consecutive
iterations, and the time step for integration of the equations of motion in MD was 0.5~fs.

The Z~method was developed by Belonoshko and collaborators. It was first employed with empirical
potentials~\cite{Belonoshko:2006p287}, and more recently in combination with first-principles
methods~\cite{Koci:2007}.
This method relies on the observation that, in the limit of superheating, i.e. when the solid is heated to the
maximum temperature that it can sustain as a metastable phase without transforming into the liquid, its
internal energy is equal to that of the liquid phase at the thermodynamic melting temperature. Thus, the
problem of determining the melting temperature can be rephrased into the the problem of determining
the temperature of the superheating limit. Then, running a microcanonical simulation of the perfect solid
at this temperature should result in the liquid phase at the equilibrium melting temperature.
We have applied this technique to calculate the
melting line of hydrogen using systems consisting of 48~molecules for the molecular phases ({\em po-hcp,
Cmca-4, Cmca-12\/}), and 128 atoms for the atomic Cs-IV phase, employing simulation times of at least 10~ps in
each case. In these simulations the Brillouin zone was sampled a $4 \times 4\times 4$~{\em k\/}-point grid, and,
like for the coexistence simulations, the time step used was 0.5~fs.

Our calculated
melting line for the {\em d-hcp\/} structure provides a higher melting line than that predicted in
previous theoretical work~\cite{Bonev:2004p296} in the range 30--200~GPa, but nevertheless in good agreement with
experimental measurements~\cite{Eremets:2009p293,Subramanian:2011}.
Moreover, we find a melting temperature for the
{\em po-hcp\/} phase of 562~K at 300~GPa, which is similar to that obtained recently~\cite{Morales:2009p323}
(550~K at 290~GPa) using free energy calculations combined with quantum Monte-Carlo methods. A further
test of the reliability of our results is the good agreement that we obtain between the melting temperatures
derived from the coexistence calculations and those resulting from the Z-method simulations
(e.g. the simulated melting temperatures of Cs-IV phase at 400 GPa are 350 K and 362 K
by using Z-method and two-phase coexistence, respectively).

Recently, phase IV has been found at room temperature and pressures above
220~GPa~\cite{Eremets:2011,Howie:2012}.
This new phase has been interpreted as a mixed structure that includes two different molecular
layers~\cite{Pickard:2012} and a partially ordered {\em hcp\/} ({\em po-hcp\/}) structure~\cite{Liu:2012p1024}.
Previous theoretical studies~\cite{Liu:2012p1024} by the present authors suggest that the {\em Cmca\/}--12 structure is less stable
than the {\em Cmca\/}--4 above 250~GPa, as the former has a higher zero-point energy than the latter.
In view of this, we have calculated the melting line of the {\em po-hcp, Cmca\/}--12 and {\em Cmca\/}--4
phases in the range 200--500~GPa using the {\em ab initio\/} Z-method. While determining the melting
temperature of the {\em Cmca\/}--4  and {\em Cmca\/}--12 phases at 300 and 350~GPa, we found that these
phases transform into the {\em po-hcp\/} structure, which indicates that the latter structure is more stable
than the other molecular structures at temperatures close to melting in this range of pressures. This result also
indicates that phase {\em Cmca\/}--12 probably has no stability field in the phase diagram of hydrogen,
which is in good agreement with our previous calculations~\cite{Liu:2012p1024}. Note that the
melting temperature of the {\em po-hcp\/} (795~K) and {\em d-hcp\/} (815~K) structures is very similar at 200~GPa,
and that the melting line of the {\em po-hcp\/} phase almost coincides with the extrapolation of that of the {\em d-hcp\/}
phase. This is consistent with the recent experimental observation that the {\em d-hcp\/} structure becomes partially ordered,  transforming
into the {\em po-hcp\/} structure at room temperature and $\sim$220~GPa~\cite{Eremets:2011,Howie:2012}.

\begin{figure}[!t]
\begin{center}
\epsfxsize=8cm
\epsffile{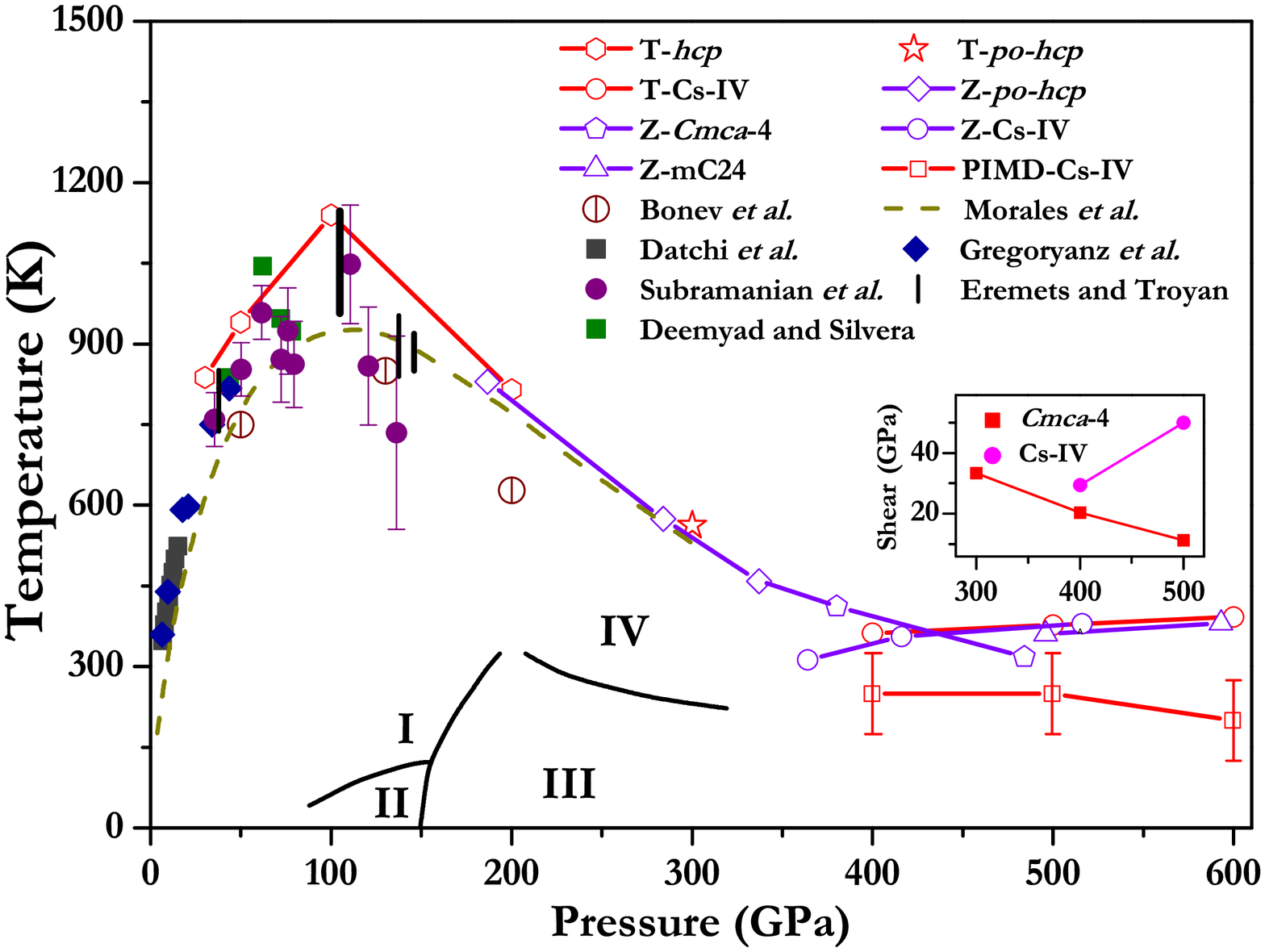}
\end{center}
\caption{(Color online) Phase diagram of hydrogen. The red and violet lines indicate the melting line obtained
from our two-phase coexistence and Z-method simulations, where T and Z mean two-phase coexistence and Z-method, respectively. The solid symbols are melting data as obtained
from previous experimental work~\cite{Deemyad:2008p505,Subramanian:2011,Datchi:2000,Gregoryanz:2003}. Theoretical melting data from
refs.~\cite{Bonev:2004p296,Morales:2009p323} is represented by open symbols. The boundaries between
phases I-II-III-IV at low temperatures are taken from
refs.~\cite{Mazin:1997,Silvera:1981,Lorenzana:1989,Howieprb:2012}. Inset shows the calculated shear modulus of {\em Cmca\/}--4 and {\em Cs\/}--IV structures as a function of pressure at 300K.
}
\label{fig:phase}
\end{figure}

At higher pressures hydrogen is expected to transform into an atomic phase. Our previous studies~\cite{HLiu:2012} have predicted
the existence of a quasi-molecular {\em mC24\/} phase in the pressure range lying between the stability fields of
the molecular {\em Cmca\/}--4 and the atomic {\em Cs\/}-IV phases. The {\em mC24\/} structure has two
different bond lengths ($\sim$0.86 and 0.9~\AA\ at 500~GPa) which are longer than that present in the
molecular {\em Cmca\/} phase (0.78~\AA\ at 400~GPa). The {\em mC24\/} phase will dissociate into the atomic
phase upon temperature increase (at $\sim$350~K), due to thermal fluctuations. Thus, in order to characterize
the melting behavior of hydrogen at pressures in the range 350--600~GPa, we have calculated the melting
temperatures of the {\em mC24\/} and {\em Cs\/}-IV phases by employing both the two-phase coexistence and Z-methods.
We find that the intersection of the melting lines of the molecular and atomic phases takes place
at $\sim$432~GPa and 367~K (see Fig.~\ref{fig:phase}), and that the melting temperature of the {\em mC24\/} structure is
systematically lower than that of the {\em Cs\/}--IV phase, indicating that the latter is more stable at temperatures
close to melting.

The most intriguing observation to be extracted from our melting temperature calculations is that the phase
diagram of hydrogen should have a second extremum along the melting line. The first extremum is the
maximum originally reported by Bonev and coworkers~\cite{Bonev:2004p296} and later
confirmed by experimental measurements~\cite{Deemyad:2008p505,Eremets:2009p293,Subramanian:2011}. The second extremum
we find here occurs at the
intersection of the melting lines of the molecular and atomic phases ($\sim$432~GPa, 367~K, see Fig.~\ref{fig:phase}) and
is a minimum, rather than a maximum. Although the molecular phase has a melting line with a negative
slope at pressures above $\sim100$~GPa, we find the atomic {\em Cs\/}--IV phase to have a melting line
with a shallow, but nevertheless positive slope above 360~GPa. Therefore, the intersection of these two melting lines occurs at a minimum.
Thus it is seen that the melting behaviors of the molecular and atomic phases are qualitatively different. In order
to further assert this difference in melting behavior, we have calculated the finite temperature shear modulus
of the {{\em Cmca\/}--4 and {\em Cs\/}-IV phases at 300~K at pressures between 300 and 500~GPa.
According to the Born melting criterion~\cite{born:1939}, melting occurs when the temperature is such that the
shear modulus of the solid phase reduces to zero. Thus we expect to observe a strong correlation between
the behavior of the shear modulus and that of the melting temperature as a function of pressure. The finite
temperature shear modulus was calculated by performing {\em ab initio} MD simulations of suitably deformed supercells of
each phase containing 96 (in the case of the {\em Cmca\/}--4 structure) or 128 atoms (for the {\em Cs\/}--IV
structure). Each run consisted of a total of 4000 steps with a time step of 0.5~fs, and the stress components were averaged over the last 2000 steps. In the case of the {\em Cmca\/}--4 phase we also run the simulation at 400~GPa
for a total of 8000 time steps, but the averaged stress components were identical to those obtained in the shorter
run. The calculated shear moduli for these phases  at 300 K are shown in the inset of Fig.~\ref{fig:phase}, where it can be seen that the
moduli of the molecular phases decrease with increasing pressure, in consonance with the behavior of their
corresponding melting lines. The shear modulus of the {\em Cs\/}-IV atomic phase, in contrast, displays the
opposite trend, which is again in accordance with its melting behavior.

Up to this point all our results have been based on classical simulations. However, hydrogen being the lightest of elements, and
the fact that the melting temperatures predicted at high pressures are relatively low, make it necessary to consider the
possible influence of quantum effects on the melting temperature. To address this issue in the particular case of the
{\em Cs\/}-IV atomic phase, we have employed path integral simulations~\cite{Marx:1996} combined with the hysteresis method~\cite{Luo:2004}
to determine the melting temperature. The hysteresis method, similar in spirit to the Z-method discussed earlier, has been
shown to be capable of producing accurate estimations of the melting temperature~\cite{Luo:2004,Bouchet:2009}.
We employed this method with a 72-atom super-cell of the {\em Cs\/}-IV structure with 8 beads in the quantum ring polymer
(doubling the number of beads to 16 did not affect the resulting melting temperature at 400~GPa),
and using $5\times 5\times 3$~{\em k\/}-points for each bead.
The resulting melting temperature at 400~GPa is $\sim$250~K, which is lower than the temperature obtained from
classical molecular dynamics (350~K), confirming the importance of quantum effects in this system.
Furthermore, it can be seen in Fig.~(\ref{fig:phase}) that the melting line including quantum effects for the {\em Cs\/}-IV
is nearly flat in the range 400-500~GPa, but seems to regain a negative slope above 500~GPa.
This could indicate the presence of a second local maximum in the melting line
of hydrogen, followed by the re-establishment of reentrant behavior at pressures in the range 500-600~GPa.
The existence of a negative slope in the melting line in this range of pressures could again be taken as a possible
indication of the existence of a quantum liquid phase at high pressure, but according to our results we do not
expect such a phase, if it exists, to be found at pressures below 600~GPa, the maximum pressure considered in this
study. In this respect, we note that a very recent study~\cite{chen:2012} employing path integral simulations has
reported melting temperatures of the order of 50~K from 900~GPa onwards.

In summary, we have calculated the melting line of hydrogen up to 600~GPa using first principles molecular
dynamics methods, considering melting from the {\em d-hcp\/}, {\em po-hcp\/}, {\em Cmca\/}--4,
{\em Cmca\/}--12 molecular phases and the {\em Cs\/}-IV atomic phase. The obtained classical melting line of
molecular hydrogen is in good agreement with previous experimental and theoretical results. We predict,
however, that the melting line reaches a minimum at a pressure of $\sim$432~GPa, coinciding with the
point at which the melting lines of the {\em Cmca\/}--4 and atomic {\em Cs\/}--IV phases cross, and that the
classical melting line retains a positive (though shallow) slope at least up to 600~GPa. Quantum effects incorporated
approximately via Feynman path integral simulations result in a downward shift of the melting line by as much
as $\sim$100~K, and hint at the possibility of a secondary maximum along the melting line in the region 500 to 600~GPa.
These results suggest that the proposed
low-temperature quantum liquid phase, if it exists at all, will not be found at pressures below 600~GPa.

The authors acknowledge funding support from the China 973 Program under Grant No. 2011CB808200, the National Natural Science Foundation of China under Grants No. 11274136, No. 11025418, and No. 91022029, 2012 Changjiang Scholar of the Ministry of Education, the open project of key lab of computational physics in Beijing Computing Physics and Applied Mathematics, and Changjiang Scholar and Innovative Research Team in University (Grant No. IRT1132). The work of ERH is funded by the Spanish Research and Innovation Office through project No. FIS2012-31713. The work of JY is supported from the Foundation for Development of Science and Technology of China Academy of Engineering Physics (Grant No. 2011A0102007). Part of the calculations were performed in the high performance computing center of Jilin University, and at the HECToR UK National Supercomputer Service. The authors would like to thank Prof. D.~Alf\`{e} for helpful discussions.

\end{document}